\documentclass[a4paper,fleqn,usenatbib]{mnras}
\usepackage{newtxtext,newtxmath}
\usepackage[T1]{fontenc}
\usepackage{ae,aecompl}
\usepackage{graphicx}	
\usepackage{amsmath}	
\usepackage{amssymb}	
\title[Multi-wavelength view of 1ES 1218+304]{Long term multi-wavelength view of the blazar 1ES 1218+304}

\author[K. K. Singh et al.]{
K. K. Singh,$^{1,2}$\thanks{E-mail: kksastro@barc.gov.in (KKS)}
B. Bisschoff,$^{1}$
B. van Soelen,$^{1}$
A. Tolamatti,$^{2}$
J. P. Marais,$^{1}$
P. J. Meintjes$^{1}$
\\
$^{1}$Physics Department, University of the Free State, Bloemfontein, 9300, South Africa\\
$^{2}$Astrophysical Sciences Division, Bhabha Atomic Research Centre, Mumbai, 400085, India
}

\date{Accepted XXX. Received YYY; in original form ZZZ}

\pubyear{2019}

\begin{document}
\label{firstpage}
\pagerange{\pageref{firstpage}--\pageref{lastpage}}
\maketitle
\begin{abstract}
In this work, we present a multi-wavelength study of the blazar 1ES 1218+304 using near simultaneous observations over 
10 years during the period September 1, 2008 to August 31, 2018 (MJD 54710-58361). We have analyzed data from 
\emph{Swift}-UVOT, \emph{Swift}-XRT and \emph{Fermi}-LAT to study the long term behaviour of 1ES 1218+304 in different 
energy bands over the last decade. We have also used the archival data from OVRO, MAXI and \emph{Swift}-BAT available 
during the above period. The near simultaneous data on 1ES 1218+304 suggest that the long term multi-wavelength emission 
from the source is steady and does not show any significant change in the source activity. The optical/UV fluxes are found 
to be dominated by the host galaxy emission and can be modelled using the $PEGASE$ code. However, the time averaged X-ray 
and $\gamma$-ray emisions from the source are reproduced using a single zone leptonic model with log-parabolic distribution 
for the radiating particles. The intrinsic very high energy $\gamma$-ray emission during a low activity state of the source is 
broadly consistent with the predictions of the leptonic model for blazars. We have investigated the physical properties of the 
jet and the mass of the super massive black hole at the center of the host galaxy using long term X-ray observations from the 
\emph{Swift}-XRT which is in agreement with the value derived using blackbody approximation of the host galaxy. We also discuss 
the extreme nature of the source on the basis of X-ray and $\gamma$-ray observations.
\end{abstract}



\section{Introduction}
Blazars are observed to be a dominant class of extragalactic $\gamma$- ray source in the Universe.
These objects are characterized as radio-loud active galactic nuclei (AGN) having elliptical
morphology and powerful relativitic plasma jets originating from its center. The relativistic
jets in blazars are oriented along the line of sight of the observer within a few degrees 
of viewing angle \citep{Urry1995,Padovani2017}. The blazar jets are believed to be powered 
by the accretion of matter onto a rotating supermassive black hole (SMBH) at the center of 
the host galaxy \citep{Blandford1977}. The non-thermal emission from the jet dominates over the 
entire electromagnetic spectrum from radio to very high energy (VHE; E $>$ 100 GeV) $\gamma$-rays. 
The small viewing angle geometry of the blazar jets leads to the Doppler boosting and beaming of 
the non-thermal radiation emitted from the outflowing plasma. The distinctive observational 
properties of the multi-wavelength (MWL) emission from blazars such as superluminal motion, 
strong anisotropy, rapid variability and a high degree of polarization are attributed to the 
relativistic beaming effects. Blazars are generally classified in two groups namely Flat Spectrum 
Radio Quasars (FSRQs) and BL Lacertae objects (BL Lacs) on the basis of their optical 
emission/absorption line features \citep{Urry1995}. FSRQs are observed to exhibit typical quasar-like 
optical spectra with strong and broad spectral lines whereas BL Lacs have mainly featureless optical 
spectra with weak/narrow or no spectral lines. The difference in the features of the optical spectra 
of the two types of blazars is attributed to the different accretion processes onto the central  
SMBH in the host galaxy \citep{Ghisellini2009}.
\par
The broadband emission from blazars is generally described by a unique spectral energy distribution 
(SED) with two characteristic humps peaking at low and high energies (HE; E $>$ 100 MeV) respectively. 
The low energy component of the SED peaks at optical/X-ray energies and its origin is associated with 
the synchrotron emission of the relativistic electrons in the magnetic field of the jet \citep{Romero2017}. 
The origin of second hump in the blazar SED peaking at $\gamma$-ray energies is not clearly understood and 
different models have been proposed depending on the specific blazar \citep{Romero2017}. For most of the blazars, 
the observed $\gamma$-ray emission is ascribed to the leptonic models in which inverse Compton (IC) scattering 
of the low energy photons by the relativistic electrons in the jet results in the production of the 
highest energy $\gamma$-ray photons. The so called synchrotron self-Compton (SSC) model is the simplest scenario 
to explain the $\gamma$-ray emission mostly from the BL Lac type of blazars through the IC scattering of 
low energy synchrotron photons by the same population of relativistic electrons that produce synchrotron 
photons in the jet \citep{Maraschi1992,Bloom1996,Bottcher2007}. If the soft target photons for the IC scattering 
originate from regions outside the jet, the process is referred to as external Compton (EC) and this model is 
generally invoked to explain the $\gamma$-ray emission from the FSRQ type of blazars \citep{Dermer1993,Sikora1994,Yan2012}. 
Time dependent one zone SSC models have also been proposed to explain the flaring activity of a few selected 
blazars \citep{Mastichiadis2013,Singh2017}. Alternatively, hadronic models are invoked to describe 
the HE $\gamma$-ray emission from a few blazars through the synchrotron radiation of ultra-relativistic protons 
in the jet magnetic field or through the photo-pion production followed by pion decay \citep{Mannheim1993, 
Aharonian2000,Murase2012}.  However, a complete understanding of the $\gamma$--ray emission from blazars 
remains to be ascertained in high energy astrophysics.
\par
In this paper, we use long term multi-wavelength (MWL) data available for the blazar 1ES 1218+304 to explore the properties of 
the broadband emission from the source. The important observational features of the source are 
briefly described in Section 2. In Section 3, we discuss the MWL observations and data analysis procedures followed for 
different instruments in the present study. Results derived from the long term MWL data are discussed 
in Section 4. Finally, the important findings of this work are summarized in Section 5. We have used  the 
$\Lambda$CDM model which provides the simplest description of the present cosmological observations with the model 
parameters H$_0$ = 70 km s$^{-1}$ Mpc$^{-1}$, $\Omega_m$ = 0.27 and $\Omega_{\Lambda} $= 0.73.

\section{1ES 1218+304}
The blazar 1ES 1218+304 first appeared in the catalogue of 3235 radio sources observed at 408 MHz with the Bologna Northern 
Cross telescope (B2 survey) in 1970 \citep{Colla1970}. In 1978, the \emph{Ariel V} catalogue of 105 high galactic latitude sources 
reported this object as a new X-ray source 2A 1219+305 with irregular variability using the complete sky monitoring (2A survey) 
in the energy range 2-18 keV \citep{Cooke1978}. After its discovery as an unidentified X-ray source in the 2A catalogue, 
this source was identified as a BL Lac object using radio and optical observations in 1979 \citep{Wilson1979}. 
Photometry revealed a break in the optical-infrared spectral flux distribution of the source \citep{Ledden1981}. 
The broadband X-ray spectra of 1ES 1218+304 observed with the \emph{Einstein Observatory} in the energy range 0.5-20 keV 
confirmed the presence of an absorption feature at an energy of $\sim$ 0.65 keV and the spectrum was found to be 
well described by a power law with Galactic gas absortion having a column density of 1.78$\times$10$^{20}$ cm$^{-2}$ 
\citep{Madejski1991}. The faint radio images of the blazar 1ES 1218+304 at 20 cm were produced by the FIRST survey using 
the observations during April-May 1993 \citep{Becker1995}. The redshift $z$ = 0.182 for 1ES 1218+304 (PG 1218+304) was 
determined in 1997 using the spectroscopic measurements of the host galaxy \citep{Bade1998}. In 2001, observations with the 
\emph{XMM-Newton} suggested that no broad absorption features are present in the  X-ray spectra of 1ES 1218+304 and the earlier 
feature observed at 0.65 keV is transient \citep{Blustin2004}. On the basis of the previous observations in radio, optical 
and X-ray energy bands, 1ES 1218+304 was predicted to be a TeV BL Lac candidate with an estimated integral flux of 
6.7$\times$10$^{-12}$ ph~cm$^{-2}$~s$^{-1}$ above 0.3 TeV energy in 2002 \citep{Costamante2002}. The MAGIC telescope 
discovered the first VHE $\gamma$-ray signal from 1ES 1218+304 with a 6.4$\sigma$ significance above an energy 
threshold of $\sim$ 0.12 TeV in 2005 \citep{Albert2006}. The integral flux above 0.35 TeV was found to be a factor two 
below the upper limit (8.3$\times$10$^{-12}$ ph~cm$^{-2}$~s$^{-1}$) estimated from Whipple observations during 1995-2000 
\citep{Horan2004}. The HEGRA telescopes also reported an upper limit on the integral flux as 
2.67$\times$10$^{-12}$ ph~cm$^{-2}$~s$^{-1}$ above 0.84 TeV using the VHE observations during 1996-2002 \citep{Aharonian2004}. 
After the discovery of VHE $\gamma$-ray emission from 1ES 1218+304 by the MAGIC telescope, the source was the target of a 
HESS observation campaign in 2006 and the upper limit on the integral flux above 1.0 TeV was estimated to be 
3.9$\times$10$^{-12}$ ph~cm$^{-2}$~s$^{-1}$ \citep{Aharonian2008}. This upper limit was a factor six above the integral 
flux estimated from the extrapolation of the MAGIC spectrum. An upper limit of 7.2$\times$10$^{-11}$ ph~cm$^{-2}$~s$^{-1}$ on 
the integral flux above 0.155 TeV was also estimated from the STACEE observations between 2006 and 2007 \citep{Mueller2011}. 
However, the VERITAS telescopes detected VHE $\gamma$-ray emission from 1ES 1218+304 with a statistical significance of 
10.4$\sigma$ in 2007 confirming the discovery by the MAGIC collaboration and the integral flux above 0.2 TeV was estimated to 
be (12.2$\pm$2.6)$\times$10$^{-12}$ ph~cm$^{-2}$~s$^{-1}$ \citep{Acciari2009}. The first evidence for the variability in 
VHE $\gamma$-ray emission from the blazar 1ES 1218+304 was detected by the VERITAS telescope during the high activity of 
the source in 2009 \citep{Acciari2010}. The average integral flux above 0.2 TeV was found to be (18.4$\pm$0.9)
$\times$10$^{-12}$ ph~cm$^{-2}$~s$^{-1}$ during this period. The highest VHE $\gamma$-ray flux was observed to be 
$\sim$ 60$\times$10$^{-12}$ ph~cm$^{-2}$~s$^{-1}$ above 0.2 TeV during the flaring activity of the source in 2009. 
The TACTIC telescope monitored this source in 2013 and reported an upper limit of 3.41$\times$10$^{-12}$ ph~cm$^{-2}$~s$^{-1}$ 
above 1.1 TeV threshold energy \citep{Singh2015}. The \emph{Fermi}-LAT (Large Area Telescoope) is continuously monitoring the 
HE $\gamma$-ray emission from 1ES 1218+304 and has reported this source as one of hardest spectrum blazar above 0.1 GeV in its 
successive catalogs \citep{Nolan2012,Acero2015,Ajello2017}.

\section{MWL Observations and data analysis}

\subsection{High energy $\gamma$- ray}
The Large Area Telescope (LAT) on board  the \emph{Fermi Gamma-ray Space Telescope} satellite (\emph{Fermi}-LAT) 
provides HE observations of the blazars in the energy range from 30 MeV to more than 500 GeV \citep{Atwood2009}. 
We have used 10 years of Pass 8 data from the \emph{Fermi}-LAT observations of the blazar 1ES 1218+304 between 
September 1, 2008 and August 31, 2018 (MJD 54710-58361) in the energy range of 0.1-300 GeV. The reprocessed Pass 8 
data sets with P8R2$\_$SOURCE$\_$V6 instrument response functions have been downloaded from the publicly available 
\emph{Fermi}-LAT data server\footnote{https://fermi.gsfc.nasa.gov/cgi-bin/ssc/LAT/LATDataQuery}. We have analyzed 
data following the standard procedures implemented in the \emph{Fermi} ScienceTools software package version \emph{v11r5p3}.
Only $SOURCE$ class events ($evclass=128$ and $evtype=3$) within a circular region of interest (ROI) of 10$^\circ$ radius centered 
at the position of 1ES 1218+304 with good time intervals and maximum zenith angle $\le$ 90$^\circ$ were selected for the analysis. 
We have included all 3FGL (Third \emph{Fermi}-LAT source catalogue) point sources \citep{Acero2015} within 20$^\circ$ from the ROI 
center as well as Galactic diffuse (gll$\_$iem$\_$vo6.fit) and extragalactic isotropic (iso$\_$P8R2$\_$SOURCE$\_$V6$\_$v06.txt) 
emission templates in the source model file for the background subtraction. The spectral shape of all the sources in the model
file are the same as reported in the 3FGL catalogue and the associated spectral parameters have been kept free to vary during 
the unbinned likelihood analysis using \emph{gtlike}. The target source 1ES 1218+304 is modelled using a power law distribution 
with normalization and spectral index allowed to vary during the fitting. The statistical significance of the HE $\gamma$-ray signal 
in the energy range 0.1-300 GeV is determined from the maximum likelihood ratio test statistic (the statistical significance 
is estimated from the square root of the value of test statistic) defined in \citep{Mattox1996}. We have derived the integral flux 
in the energy range 0.1-300 GeV averaged over 60 days for 1ES 1218+304 so that the individual flux points in the light curve have TS 
$\ge$ 15 ($\sim$ 4$\sigma$ significance).

\subsection{Hard X-ray}
We have used the archival data from hard X-ray observations of the blazar 1ES 1218+304 in two enegry bands 15-50 keV and 10-20 keV 
available during the period considered in this study. The Burst Alert Telescope (BAT) on board the  \emph{Neil Gehrels Swift Observatory}
(\emph{Swift-BAT}) provides daily light curves of blazars in the energy range 15-50 keV using transient observations \citep{Krimm2013} 
and the data are available to the public online\footnote{https://swift.gsfc.nasa.gov/results/transients}. The MAXI (Monitor of All sky 
X-ray Image) (MAXI) telescope on board the JEM satellite has started regular observations since August 2009 \citep{Matsuoka2009}. 
We have used the MAXI on-demand process\footnote{http://maxi.riken.jp/mxondem/} to obtain the daily light curve of the 
blazar 1ES 1218+304 in the energy range 10-20 keV. We have considered only 2$\sigma$ photon flux measurements to convert into the 
corresponding energy flux values using the appropriate mean energies for both the data sets from \emph{Swift}-BAT and MAXI observations.  

\subsection{Soft X-ray}
The X-Ray Telescope (XRT) on board the \emph{Neil Gehrels Swift Observatory} (\emph{Swift}-XRT) provides observations of blazars at 
soft X-rays in the energy range 0.2-10 keV \citep{Burrows2005}. We have used the XRT data collected in photon counting (PC) and 
window timing (WT) modes for the blazar 1ES 1218+304 during the period September 1, 2008 to August 31, 2018 from the \emph{Swift}
multi-wavelength observation program\footnote{https://www.swift.psu.edu/monitoring}. The XRT observations with exposure time more 
than 1 ks have been used for analysis using the standard XRTDAS tools distributed within the \emph{HEAsoft} package 
(v6.24)\footnote{https://heasarc.gsfc.nasa.gov/lheasoft/download.html}. We have used the \emph{xrtpipeline} script with 
recent calibration files (CALDB) to analyse the data in the energy range 0.3-10 keV. The source and background spectra are 
produced using \emph{xselect} tool. The ancillary response files (ARFs) have been generated using \emph{xrtmkarf} task after 
applying the corrections due to the point spread function and CCD defects. All the observations are binned to have a minimum of 
20 counts per spectral bin using \emph{grppha} utility. The spectral fitting is performed using a power law model with Galactic absorption 
using \emph{xspec} model $phabs \times zpow$ for source redshift $z$ = 0.182. The equivalent line-of-sight neutral hydrogen column 
density is fixed to the value $ n_H = 1.99 \times 10^{20} ~cm^{-2}$ obtained from the LAB (Leiden/Argentine/Bonn) Survey of 
Galactic HI \citep{Kalberla2005}.

\subsection{Ultra-violet/Optical}
The UV/Optical Telescope (UVOT) on board the \emph{Neil Gehrels Swift Observatory} (\emph{Swift-UVOT}) provides observations of 
blazars in the wavelength range 160-600 nm \citep{Roming2005}. The data for individual sources are made available in three 
ultraviolet (W1,M2,W2) and three optical (V,B,U) filters. We have analyzed the \emph{Swift}-UVOT data available from the daily 
observations of the blazar 1ES 1218+304 during September 1, 2008 to August 31, 2018 using the online UVOT Interactive Analysis 
tools\footnote{https://www.ssdc.asi.it/mmia/index.php?mission=swiftmastr}, which are based on the standard software tasks included 
in the \emph{HEAsoft} package (v6.23). The data were analyzed using \emph{uvotdetect} task and 20170922 version of CALDB . 
We have extracted the source counts from a circular region of 5$^{\prime\prime}$ radius centered at the source position, while background 
counts are derived from an annulus region with an inner radius of 27$^{\prime\prime}$ and an outer radius of 35$^{\prime\prime}$, 
centered at the source location. The magnitude at the source position is dereddened using the value of E(B-V) equal to 0.0172 according to 
\citep{Schlegel1998,Schlafly2011} and  mean galactic extinction curve reported in \citep{Fitzpatrick1999}. The dereddened magnitudes 
of the source from different filters are then converted into corresponding energy flux points.

\subsection{Radio}  
We have obtained the radio data on 1ES 1218+304 (J1217+3007) available between September 1, 2008 and August 31, 2018 from the OVRO 
(Owens Valley Radio Observatory) 40 M blazar monitoring program \citep{Richards2011}. The OVRO 40 M telescope has been 
regularly monitoring large number of blazars since 2008 and provides daily light curves for individual sources. The long 
term radio data at 15 GHz is available online to the public from the monitoring of selected \emph{Fermi} 
blazars\footnote{http://www.astro.caltech.edu/ovroblazars/data.php?page=data}. The details of OVRO observation program and data anlayis 
procedure are described in \citep{Richards2011}.

\section{Results and Discussion}

\subsection{Long term behaviour of 1ES 1218+304}
\begin{figure}
\centering
\includegraphics[width=1.0\columnwidth]{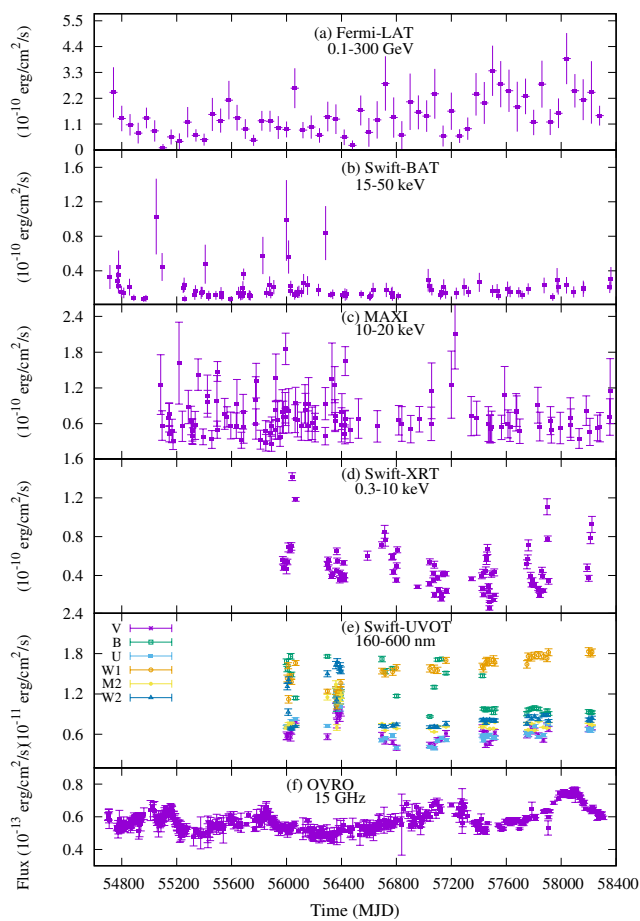}
\caption{Long term (10 years) multi-wavelength light curves for the blazar 1ES 1218+304 during the period September 1, 2008 and 
	August 31, 2018 (MJD 54710-58361). The flux points shown in the \emph{Fermi-LAT} light curve (a) are averaged over 60 days,
	The lower energy light curves from X-ray to radio (b-f) involve daily flux points measured from the different instruments. 
	The outliers visible in the BAT light curve (b) with relatively large error bars can be attributed to the fluctuations 
	in the emission from the source and uncertainty in the measurement due to moderate sensitivity of the instrument to detect 
	faint X-ray sources and less exposure time during the particular observation.}
\label{fig:Fig1}
\end{figure}

The MWL light curves of the blazar 1ES 1218+304 over a decade from September 1, 2008 to August 31, 2018 (MJD 54710-58361) shown in 
Fig.~\ref{fig:Fig1}(a-f), characterize the long term emission behaviour of the source in different energy bands. 
In Figure \ref{fig:Fig1}(a), we have reported 60 days binned flux points in the energy range 0.1-300 GeV from the \emph{Fermi}-LAT observations. 
All the flux points in the \emph{Fermi}-LAT light curve of the source have statistical significance more than 4$\sigma$. The two months 
averaged light curve from the long term \emph{Fermi}-LAT observations suggests moderate HE $\gamma$-ray emission from the blazar and indicates 
a low activity state of the source. The average flux level in the energy range 0.1-300 GeV is found to be (8.03$\pm$0.81)$\times$10$^{-11}$ 
erg~cm$^{-2}$~s$^{-1}$ which is approximately eight times higher than the upper limit derived from the EGRET observations of the source 
above 0.1 GeV during 1991-1995 \citep{Hartman1999}. The hard X-ray emissions from the blazar 1ES 1218+304 during the above period observed with 
\emph{Swift}-BAT and MAXI in the energy bands 15-50 keV and 10-20 keV are shown Figure \ref{fig:Fig1}(b) \& (c) respectively. The daily light curves 
from the \emph{Swift}-BAT and MAXI observations in \ref{fig:Fig1}(b) \& (c) respectively include only 2$\sigma$ flux points. The average X-ray 
flux levels in the energy bands 15-50 keV and 10-20 keV from the source are observed to be 
(1.37$\pm$0.16)$\times$10$^{-11}$ erg~cm$^{-2}$~s$^{-1}$ and (6.08$\pm$0.26)$\times$10$^{-11}$ erg~cm$^{-2}$~s$^{-1}$ respectively. 
It is observed from the \emph{Swift}-BAT light curve that most of the flux measurements are consistent with the average emission level 
of the source in a low activity state. Near simultaneous soft X-ray emissions observed with the \emph{Swift}-XRT in the energy range 0.3-10 keV 
are shown in Figure \ref{fig:Fig1}(d). The daily averaged flux points in the \emph{Swift}-XRT light curve suggest an average flux level of 
(3.65$\pm$0.23)$\times$10$^{-11}$ erg~cm$^{-2}$~s$^{-1}$ which is found to be in agreement with the \emph{XMM-Newton} measurements of 
1ES 1218+304 in the energy range 2-10 keV in 2001 \citep{Blustin2004}. It is important here to mention that one day \emph{NuSTAR} observation 
of 1ES 1218+304 in 2015 found a hint of intraday variability in the X-ray emission of the source in the energy range 3-79 keV and the 
average flux level in this energy band was estimated to be (1.19$\pm$0.03)$\times$10$^{-11}$ erg~cm$^{-2}$~s$^{-1}$ \citep{Pandey2018}. 
The daily light curves from the \emph{Swift}-UVOT observations in three optical (V,B,U) and three UV (W1,M2,W2) bands are reported in 
Figure \ref{fig:Fig1}(e). The long term UV/optical emissions in the wavelength range 160-600 nm have an average flux level 
$\sim$ 10$^{-11}$ erg~cm$^{-2}$~s$^{-1}$. In 2010, optical observations of the blazar 1ES 1218+304 in the R band on three nights suggested 
a hint of variability on one night \citep{Krishna2011}. A clear fading by 0.1 magnitude in the R band was also detected between the 
first two nights separated by 10 days. The daily radio light curve at 15 GHz from the OVRO telescope depicted in Figure \ref{fig:Fig1}(f) 
gives an average flux level of (5.73$\pm$0.03)$\times$10$^{-14}$ erg~cm$^{-2}$~s$^{-1}$ during the 10 years of observations of the source. 
The long term MWL light curves of the blazar 1ES 1218+304 suggest that the activity of the source in different energy bands over 
10 years can be characterized as the low emission state with some fluctuations in X-ray energy bands. However, no significant change in the 
broadband emission activity of the source is observed in this study.
\subsection{Variability}
\begin{figure}
\centering	
\includegraphics[width=0.72\columnwidth,angle=-90]{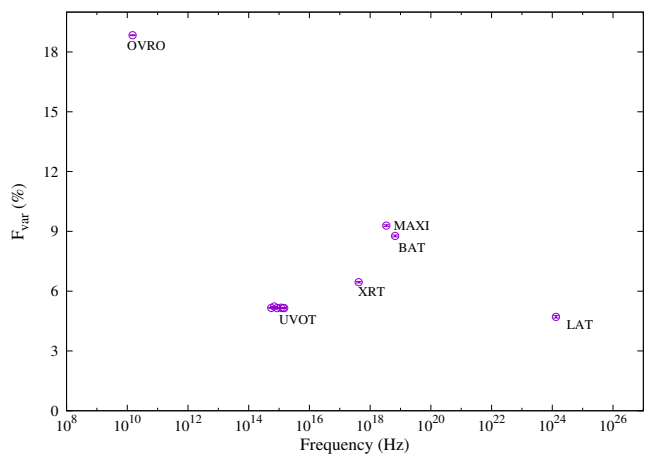}
\caption{Fractional variability amplitude ($F_{var}$) parameter for the blazar 1ES 1218+304 from radio to HE $\gamma$-rays using 
	10 years of observations during September 1, 2008 - August 31, 2018 (MJD 54710-58361) with different instruments.}
\label{fig:Fig2}
\end{figure}

The flux variability in a given light curve is a convolution of the intrinsic physical processes in the source and statistical 
uncertainty of the measurement. In order to characterize the presence of intrinsic variability in the MWL light curves of the source, 
we have estimated the fractional variability amplitude ($F_{var}$) parameter in different energy bands because it is corrected for 
the measurement noise. The variability amplitude characterized by $F_{var}$ is calculated from the intrinsic excess variance of a light curve 
after subtracting out the contribution of additional variance due to the uncertainty in the individual flux measurements. 
The $F_{var}$ is defined as \citep{Vaughan2003}  
\begin{equation}\label{Fvar}
	F_{var}=\sqrt{\frac{S^2 -E^2}{F^2}}
\end{equation}
and the formal error in $F_{var}$ is given by \citep{Vaughan2003}
\begin{equation}
	\Delta F_{var}=\sqrt{\left(\sqrt{\frac{1}{2N}}\frac{E^2}{F^2F_{var}}\right)^2+\left(\sqrt{\frac{E^2}{N}}\frac{1}{F}\right)^2}
\end{equation}  
where $S^2$ is the variance of the light curve, $F$ is the average flux, $E^2$ is the mean of the squared error in the flux measurements and 
$N$ is the number of flux points in a light curve. The intrinsic variability characterization using the $F_{var}$ parameter is important due to 
the fact that it takes into account the errors in the flux measurements. The contribution of measurement errors (Poisson noise) is  
subtracted from the variance of a light curve for the calculation of $F_{var}$. The estimated values of $F_{var}$ using Equation (\ref{Fvar})
as a function of the mean energy for different instruments are shown in Figure \ref{fig:Fig2}. It is clear from Figure \ref{fig:Fig2} that 
the radio emission at 15 GHz exhibits more variability than higher energy light curves of the source. The optical/UV and HE $\gamma$-ray 
emissions show very small intrinsic variations whereas the hard X-ray light curves from MAXI and \emph{Swift}-BAT have relatively more 
intrinsic variability.  The values of $F_{var}$ strongly depend on the time-binning of the light curves. A light curve with longer 
time bins can give give smaller values of $F_{var}$ because large time bins smooth out the variability. Therefore, the lowest value of 
$F_{var}$ for the \emph{Fermi}-LAT light curve can be attributed to the 60 days time binning. The sampling of the 
daily radio light curves is better than other daily light curves and hence the higher value of $F_{var}$ for radio. 
The values of $F_{var}$ in the different energy bands suggest that the long term broadband emission from the blazar 
1ES 1218+304 is nearly steady with small amplitude fluctuations and does not show any signature of short term variability. 
Whereas the VERITAS telescope detected a flaring activity with variability on days timescale in the VHE band during Januray 2009 monitoring 
of the blazar 1ES 1218+304 \citep{Acciari2010}.
\subsection{X-ray Photon Index-Flux Correlation}
\begin{figure}
\centering
\includegraphics[width=0.72\columnwidth,angle=-90]{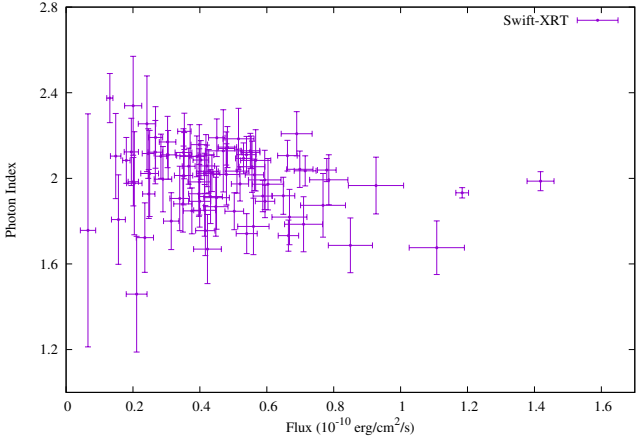}
\caption{Scatter plot for the correlation between soft X-ray flux (0.3-10 keV) and photon index of the blazar 1ES 1218+304 observed 
	 with the \emph{Swift}-XRT during September 1, 2008 - August 31, 2018 (MJD 54710-58361).}
\label{fig:Fig3}
\end{figure}

The correlation between the X-ray integral flux in the energy range 0.3-10 keV and photon index from the long term observations of 
the blazar 1ES 1218+304 using the \emph{Swift}-XRT is shown in Figure \ref{fig:Fig3}. We observe a hint for spectral hardening in the 
scatter plot with decreasing photon index as a function of integral flux. We have performed a Pearson correlation analysis to quantify 
the degree of linear correlation between the integral flux and photon index. The value of the Pearson correlation coefficient is obtained 
is $\sim$ -0.20 with a probability of null hypothesis as 0.067. This suggests that the value of Pearson correlation coefficient
for the X-ray photon index and integral flux is not very significant at the 99$\%$ confidence level. The power law photon index 
in the energy range 0.3-10 keV is found to be consistent with an average value of 1.99$\pm$0.16 and the average integral 
flux above 0.3 keV is (3.65$\pm$0.23)$\times$10$^{-11}$ erg~cm$^{-2}$~s$^{-1}$ from the long term observations of the 
blazar 1ES 1218+304 with the \emph{Swift}-XRT. The long term X-ray spectral study of 1ES 1218+304 reported by 
\cite{Wierzcholska2016} suggests that the X-ray spectrum of the source can also be described by a log-parabola model with the values 
of the spectral index and curvature parameter being $\sim$ 2.0 and $\sim$ 0.1 respectively for fixed values of Galactic absorption 
taken from different surveys. Whereas if the Galactic absoprtion is kept free, the X-ray spectrum is described by a spectral index of 
1.96$\pm$0.06 and curvature parameter of 0.22$\pm$0.08. The integral flux in the energy range 0.3-10 keV is estimated to be 
$\sim$ 3.33$\times$10$^{-11}$ erg~cm$^{-2}$~s$^{-1}$ which is consistent with the value of integral flux obtained in the present 
study of the source. Therefore, the long term spectral behaviour in the energy range 0.3-10 keV suggests that 1ES 1218+304 is a 
hard X-ray spectra blazar with a power law photon index $\sim$ 2.0.

\begin{table}
\centering	
\caption{Summary of the VHE $\gamma$-ray observations from the blazar 1ES 1218+304 between 2005 and 2018.}
\label{tab:vhetab}
\begin{tabular}{lccclc}
\hline
Epoch	&Energy Range	&$E_0$	     &$\Gamma$	&Reference\\
	&(TeV)		&(TeV)	     &          &   \\ 
\hline
2005	&0.087-0.630	&0.25	               &3.0$\pm$0.4   &MAGIC \citep{Albert2006}\\			
2007	&0.16- 1.8 	&0.50                  &3.08$\pm$0.34 &VERITAS \citep{Acciari2009}\\
2009	&0.16-1.8	&0.50                  &3.07$\pm$0.09  &VERITAS \citep{Acciari2010}\\
\hline
\end{tabular}
\end{table}

\subsection{$\gamma$-ray Spectra}
The time averaged HE $\gamma$-ray spectrum of the blazar 1ES 1218+304 is well described by a simple power law with 
a photon spectral index of 1.67$\pm$0.05 in the energy range 0.1-300 GeV from the \emph{Fermi}-LAT observations 
during during September 1, 2008 - August 31, 2018 (MJD 54710-58361). This is consistent with the value reported 
in the 3FGL catalogue based on first four years of the \emph{Fermi} observations \citep{Acero2015}. In the VHE band, statistically 
significant detections of $\gamma$-ray photons from 1ES 1218+304 at three occasions have been reported by ground based 
Cherenkov telescopes like MAGIC and VERITAS. The observed differential VHE $\gamma$-ray spectra are described by a power 
law of the form 
\begin{equation}\label{vhespec}
	\left(\frac{dN}{dE}\right) = N_0~\left(\frac{E}{E_0}\right)^{-\Gamma}
\end{equation}
where $N_0$ is the flux normalization at energy $E = E_0$ and $\Gamma$ is the observed photon spectral index. The values of 
photon spectral indices derived from the VHE $\gamma$-ray observations of 1ES 1218+304 at different epochs are summarized in 
Table \ref{tab:vhetab} from the literature. It is evident from the values of $\Gamma \sim$ 3.0 that the observed 
VHE $\gamma$-ray spectra of the blazar 1ES 1218+304 at three epochs is softer than the HE spectrum measured from the \emph{Fermi}-LAT.
\par
\begin{figure}
\centering
\includegraphics[width=0.72\columnwidth,angle=-90]{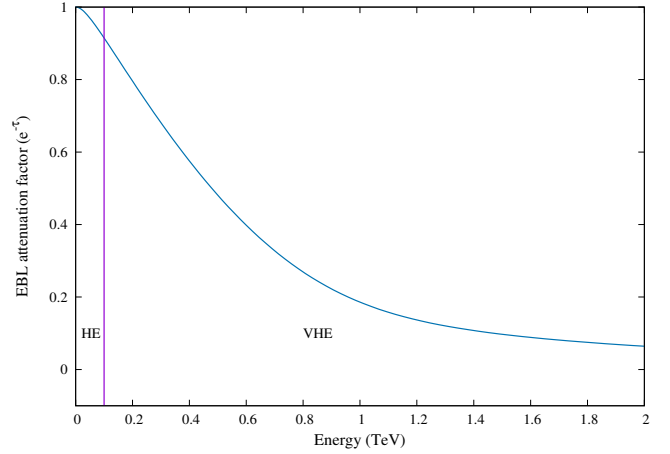}
\caption{EBL attenuation factor for $\gamma$-ray photons in the energy range 0.1 GeV - 2 TeV emitted from the blazar 
	 1ES 1218+304 at redshift $z = 0.182$ using EBL model proposed by \citet{Finke2010}. The solid vertical line 
	 separates HE and VHE regions.}
\label{fig:Fig4}
\end{figure}

The softening of the observed VHE spectra can be attributed to the absorption of TeV photons by the low energy extragalactic 
background light (EBL) due to the high redshift ($z = 0.182$) of the source. The $\gamma$-ray photons are attenuated by interacting with 
the IR/optical photons of the EBL via $e^- e^+$ pair production while propagating from source to the observer \citep{Gould1966}. 
The attenuation of photons is characterized by the optical-depth ($\tau$) which strongly depends on the density of EBL photons, observed 
$\gamma$-ray photon energy ($E$) and redshift ($z$) of the source. For a given source, the observed $\gamma$-ray flux 
($F_{obs}$) is related to the emitted flux ($F_{emi}$) by the relation
\begin{equation}\label{intspec}
	F_{obs} (E) = F_{emi} (E) ~e^{-\tau(E,~z)}
\end{equation}  
where $e^{-\tau(E,~z)}$ is referred to as the \emph{EBL attenuation factor}. In order to characterize the attenuation of $\gamma$-ray 
photons emitted from the blazar 1ES 1218+304, we have estimated the optical depth ($\tau$) in the energy range 0.1 GeV-2 TeV following 
the methodology described in \citep{Singh2014} for the density of EBL photons predicted by \citet{Finke2010}.
The corresponding \emph{EBL attenuation factor} as a function of $\gamma$-ray photon energy for redshift $z = 0.182$ is shown in 
Figure \ref{fig:Fig4}. It is evident from Figure \ref{fig:Fig4} that the EBL attenuation is negligible for HE photons whereas 
VHE photons are significantly attenuated. Therefore, observed VHE flux points from the blazar 1ES 1218+304 must be corrected 
for EBL absorption using Equation (\ref{intspec}) to predict the intrinsic VHE spectra emitted from the source. The intrinsic 
VHE spectra of 1ES 1218+304 during three epochs (Table \ref{tab:vhetab}) are found to be described by a power law with photon 
spectral indices $\Gamma \le$ 2. \citet{Acciari2009} have constrained the intrinsic VHE spectrum of the source to be harder 
than $\Gamma$= 1.86$\pm$0.37. The \emph{Fermi}-LAT observations in the energy 10 GeV - 2 TeV have also estimated  
hard power law spectrum for the blazar 1ES 1218+304 with $\Gamma$= 1.472$\pm$0.156 in the Third Catalog of Hard \emph{Fermi}-LAT 
Sources (3FHL) using the first 7 years of data \citep{Ajello2017}. These results suggest that the $\gamma$-ray spectra of 
1ES 1218+304 is a hard power law with photon spectral index $\Gamma \le$ 2. During the 2009 campaign of the source, a significant 
increase in the flux level above 0.12 TeV was detected by the VERITAS telescope with the highest flux level on the nights of 
January 30 and 31 \citep{Acciari2010}. However, no evidence for change in the VHE spectral index was found during the high 
activity state of the source. This suggests that the harder-when-brighter feature of blazars was not observed in the VHE 
$\gamma$-ray emission from the blazar 1ES 1218+304.
\subsection{Spectral Energy Distribution}
\begin{figure}
\centering
\includegraphics[width=0.72\columnwidth,angle=-90]{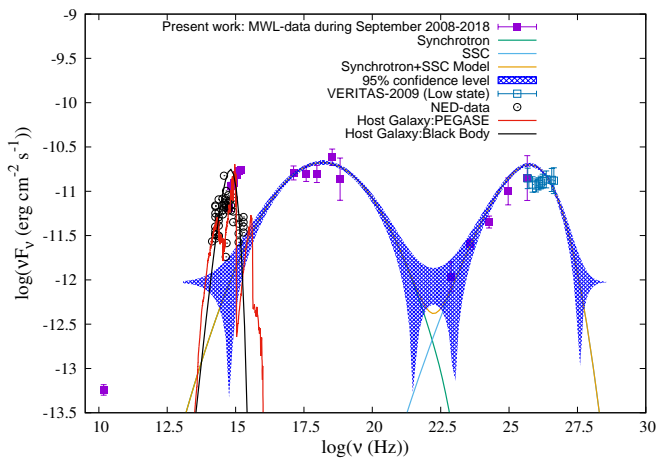}
\caption{Broadband spectral energy distribution of the blazar 1ES 1218+304 from the near simultaneous long term 
	  observations over 10 years. The data points in the present work from radio to HE $\gamma$-ray correspond to 
	  the time averaged flux measurements from OVRO, \emph{Swift}-UVOT, \emph{Swift}-XRT, MAXI, \emph{Swift}-BAT and 
	  \emph{Fermi}-LAT observations of the source during 2008-2018. The VHE flux points are from the VERITAS 
	  observations in low activity state of 1ES 1218+304 during 2009 \citep{Acciari2010} and have been corrected for EBL 
	  absorption using \citep{Finke2010} model. The solid lines depict different model curves used to reprodce 
	  the broadband emission. The blue band is the 95$\%$ confidence level for the model SED obtained from 
	  the simple one zone SSC model to include the effects of variability in different energy bands.}
\label{fig:Fig5}
\end{figure}

The time averaged broadband emission from the blazar 1ES 1218+304 in radio, optical/UV, X-ray and $\gamma$-ray bands measured from 
different instruments during September 1, 2008 - August 31, 2018 is depicted in Figure \ref{fig:Fig5}. In the VHE $\gamma$-ray band, we 
have used flux points from the VERITAS observations of the source in a low activity state during January - May 2009 \citep{Acciari2010}.
The VHE flux points from the  VERITAS observations have been corrected for EBL absorption using the model for density of low energy 
background photons proposed in \citep{Finke2010}. We use a simple one zone leptonic model to reproduce the broadband emissions from  
the blazar 1ES 1218+304. The model is fully described in \citep{Tramacere2011,Tramacere2009,Massaro2006}. In brief, the non-thermal 
emission region in the jet is assumed to be a spherical blob of radius $R$ filled with plasma of relativistic leptons (e$^{\pm}$) 
and cold hadrons (protons at rest). The emission region is entangled with a constant and uniform magnetic field $B$. The distribution 
of relativistic electrons in the emission region is described by a $log-parabolic$ function. The differential electron distribution per 
unit volume is given by 
\begin{equation}\label{eleceqn}
	n(\gamma) = K \left(\frac{\gamma}{\gamma_0}\right)^{- s - r~log~(\frac{\gamma}{\gamma_0})}; ~~~~~\gamma_{min} \le \gamma \le \gamma_{max}
\end{equation}
where $K$ is the normalization constant, $\gamma$ is the Lorentz factor of electrons (and positrons) in the blob rest frame, $\gamma_0$ is the 
turn-over Lorentz factor, $s$ and $r$ are spectral index and curvature parameter respectively. $\gamma_{min}$ and $\gamma_{max}$ are the 
Lorentz factors corresponding to the minimum and maximum energies of the electron distribution. The relativistic electrons described by 
Equation (\ref{eleceqn}) emit electromagnetic radiation through synchrotron and SSC processes in the emission region. The emission region 
is considered to be relativistically moving and therefore the emitted radiation is boosted by the Doppler factor given by  
\begin{equation}
	\delta = \frac{1}{\Gamma_j(1-\beta_j cos~\theta)}
\end{equation} 
where $\beta_j$ is bulk speed of the jet in the units of speed of light in vacuum $c$, $\Gamma_j$ is the bulk Lorentz factor and 
$\theta$ is viewing angle. The observed synchrotron and SSC fluxes are estimated using the AGN-SED simulator 
tool\footnote{http://www.isdc.unige.ch/sedtool/PROD/SED.html}. The theoretical braodband SED calculated for the blazar 1ES 1218+304 
from the model described above is shown in Figure \ref{fig:Fig5}. It is observed that the simple one zone SSC model with parameters reported 
in Table \ref{tab:sedpar} satisfactorily reproduces the broadband emission from 1ES 1218+304 except the optical/UV flux points. 
The optical/UV measurements are higher than the non-thermal emission from the jet predicted by the SSC model. 
Also, the long term observations of 1ES 1218+304 reported in the NASA/IPAC Extragalactic Database 
(NED)\footnote{https://ned.ipac.caltech.edu} indicate that the UV/optical/IR flux points are higher than the jet emission. These observations 
suggest that the stellar emission from the host galaxy of the source is dominant at optical/UV frequencies. In order to accurately account for 
the stellar emission due to the host galaxy, we have used the \emph{PEGASE.3} (Program for the study of galaxies by evolutionary synthesis) 
code\footnote{http://www2.iap.fr/pegase} to model the optical/UV fluxes \citep{Fioc2019}. The \emph{PEGASE.3} code takes into account the 
evolution of stars, gas and dust, and their radiative energy output to reliably model the evolution of the host galaxy. We have used the 
default values provided in the \emph{PEGASE.3} code to initiate the simulation. The evolution of host galaxy is modelled using three gas 
infall periods lasting 3$\times10^9$ years each representing the historic mergers during the evolution and last infall ending 10$^9$ years 
before the present epoch. Each infall initiates an intense star and dust formation epoch as well as reddenning of the host galaxy spectrum. The 
resulting spectrum at 14.4$\times10^9$ years is normalized to the mass of galaxy hosting the blazar 1ES 1218+304 and has been converted to flux 
at the redshift  $z = 0.182$. The associated mass of the host galaxy is found to be $\sim$~5.12$\times10^{10}~M_\odot$. 
The final spectral energy distribution of the host galaxy is shown in Figure \ref{fig:Fig5} along with the NED data points. 
We observe that stellar emission from the host galaxy predicted by the \emph{PEGASE.3} code dominates over the non-thermal jet emission and 
is broadly consistent with the optical/UV flux measurements from the source. We have also used a black-body approximation with a radius of 
9.7$\times$10$^{15}$ cm and a temperature of $\sim$ 9000K to reproduce the stellar emission due to the host galaxy of the blazar 1ES 1218+304.
\par
The log-parabola distribution of electrons in the emission region can be an outcome of statistical and stochastic acceleration processes 
with acceleration efficiency inversely proportional to the energy of electrons in the jet \citep{Massaro2006}. The fluctuations in the energy 
gain by the accelerating particles introduce a curvature ($r$) in the power law spectrum ($s$) and hence the broadening in the spectral shape. 
The curvature in the log-parabolic distribution is also related to the diffusion in the momentum space with curvature parameter ($r$) 
being inversely proportional to the diffusion coefficient \citep{Tramacere2011}. This leads to an inverse correlation between the 
curvature parameter and peak frequencies in the low (synchrotron) and high (IC) energy components of the broadband SED. Therefore, the 
curvature in the spectrum can be attributed to the acceleration process and radiative cooling of relativistic electrons. The radiative 
cooling due to synchrotron and IC processes moves high energy radiating particles (electrons) to lower energies. This results in the 
increased curvature at low energies and soft spectrum at high energies. \citet{Ruger2010} have fitted the contemporaneous data 
in X-ray and VHE $\gamma$-ray bands on 1ES 1218+304 using a SSC model involving an electron distribution as a power law with exponential 
cut-off. The model parameters inferred from this fitting suggest that the power spectral index $s=2.1$ can be due 
to the \emph{Fermi} acceleration process. However, this was found to be inconsistent with the hard electron distribution with 
$s=1.7$ derived from the X-ray observations during flaring activity of the source in the energy range 5-10 keV in 2006 \citep{Sato2008}. 
\citet{Archambault2014} used simultaneous X-ray and VHE observations as well as contemporaneous optical and HE data 
to model the broadband SED of 1ES 1218+304 with a single zone SSC model assuming the electron energy spectrum 
described by a broken power law. A spectral break in the electron spectrum larger than that expected from the 
radiative cooling involving synchrotron and IC processes was obtained from the SED modelling. This indicated 
the contribution of particle acceleration to the observed spectral break in the electron energy spectrum. 
\citet{Ding2017} have modelled the quasi-simultaneous broadband SED of 1ES 1218+304 using a single zone SSC model with the 
electron energy distribution as log-parabolic spectrum. Recently, the electron distribution with smooth broken power law 
and very small magnetic field ($B \sim 0.0035$G) has also been used to fit the broadband SED of 1218+304 under one zone SSC model 
\citep{Costamante2018}. A hadronic model with proton synchrotron and lepto-hadronic scenarios has also been invoked to explore 
the properties of broadband emissions from the blazar 1ES 1218+304 \citep{Cerruti2015}. Therefore, the model parameter space 
for broadband emission from 1ES 1218+304 is degenerate like many other blazars and the parameters derived in the present study 
(Table \ref{tab:sedpar}) represent one of the probable set of model parameters for this source. The set of model parameters 
obtained in this study corresponds to the lowest jet Doppler factor ($\delta \approx$ 26) as compared to the values 
derived from different SED modelling by various authors discussed above. As discussed in Section 4.2, the estimated 
values of $F_{var}$ in different energy bands suggest nearly steady broadband emission from the source during the last decade. 
However, higher values of the $F_{var}$ for a given light curve indicate flux variability, which can be associated with the spectral 
change in that energy band. This spectral change can cause variation in the SED which strongly depends on the spectrum of emitting 
particles. Therefore, the physical parameters of the model may vary significantly from the values reported in Table \ref{tab:sedpar}.
Hence, we have constructed a 95$\%$ confidence level for the model SED in Figure \ref{fig:Fig5} to include the effect of observed small 
variability in the X-ray and HE $\gamma$-ray bands. 

\begin{table}
\centering
\caption{Parameters from the broadband SED modelling of the blazar 1ES 1218+304 using a simple one zone synchrotron and SSC model 
	with log-parabolic distribution function for electrons.}
\label{tab:sedpar}	
\begin{tabular}{lccc}
\hline
Parameter		   		&Symbol      	&Value\\
\hline
Redshift                                &$z$            &0.182\\    
Size of the emission region  		&$R$		&2.2$\times$10$^{15}$ cm\\
Jet viewing angle                       &$\theta$	&1.4$^\circ$\\
Bulk Lorentz factor of jet 		&$\Gamma_j$ 	&15	\\
Doppler factor                  	&$\delta$       &26     \\
Electron spectral index                 &$s$		&1.8	\\
Curvature parameter                     &$r$            &0.5     \\
Turnover Lorentz factor                 &$\gamma_0$	&2.5$\times$10$^{4}$ \\
Maximum Lorentz factor                  &$\gamma_{max}$ &5.5$\times$10$^{7}$ \\
Minimum Lorent factor                   &$\gamma_{min}$ &6\\
Magnetic field                          &B              &0.22 G\\
Electron energy density                 &U$_e$           &0.16 erg~cm$^{-3}$\\
Proton energy density                   &$U_p$		&1.8$\times$10$^{-3}$ erg~cm$^{-3}$\\
\hline
\end{tabular}
\end{table}

\subsection{Physical Properties of Jet}
The modelling of the broadband emission using the radiative processes can be used to probe some of the physical properties of the emission region 
in the relativistic jet of the blazar. From the SED modelling of 1ES 1218+304, the jet is found to be closely aligned along the line of sight 
of the observer with viewing angle of $\sim$ 1.4$^\circ$. The emission region composing of electron-proton plasma moves relativistically along 
the jet axis with bulk Lorentz factor $\sim$ 15. In the leptonic emission scenario, the radiative output is dominated by the relativsitic electrons 
and the protons are at rest due to their larger mass in the comoving frame. Therefore the contribution of protons to the radiative output is not 
significant, but they can carry a significant fraction of jet kinetic power. The total kinetic power of the jet in the rest frame of the host galaxy 
of 1ES 1218+304 is given by \citep{Celotti1997}
\begin{equation}
	P_{jet} \approx \pi R^2 \Gamma^2_j c(U_e + U_B + U_p)	
\end{equation}
where $U_e$, $U_B$ are $U_p$ are comoving energy densities assocaited with electrons, magnetic field and cold protons respectively. 
From the best fit SED parameters reported in Table \ref{tab:sedpar}, the kinetic power of jet is found to be $\sim$ 1.24$\times$10$^{44}$ 
erg~s$^{-1}$ and the total radiative power is $\sim$ 3.2$\times$10$^{42}$ erg~s$^{-1}$. This indicates that only small fraction of 
the total jet power is converted into the radiative output. Assuming that the blazar jets are driven by the rotation and accretion 
of the SMBH at the center of the host galaxy, the jet power strongly depends on the mass of black hole \citep{Davis2011}. The correlation 
between the intrinsic X-ray luminosity in the energy range 2-10 keV ($L_{2-10~keV}$) and mass of SMBHs ($M_{BH}$) is described by 
the empirical relation \citep{Mayers2018}
\begin{equation}\label{mbh}
	M_{BH} = 2.0 \times 10^{7} \left(\frac{L_{(2-10~ \rm keV)}}{1.15 \times 10^{43}~ \rm erg~s^{-1}}\right)^{0.75}~ M_\odot
\end{equation}
where $M_\odot$ is the solar mass. Using the value of $L_{(2-10~ \rm keV)} \sim$ 1.12$\times$10$^{45}$ erg~s$^{-1}$ from the \emph{Swift}-XRT 
observations in Equation (\ref{mbh}), we get an estimate for the mass of SMBH as $\sim$ 6.0$\times$10$^8$ $M_\odot$ for the blazar 
1ES 1218+304. This is in agreement with the central black hole mass 5.6$\times$10$^8$ $M_\odot$ estimated for 1ES 1218+304 
using black body approximation for the host galaxy \citep{Ruger2010}. The Eddington luminosity for the SMBH is given by 
\begin{equation}
	L_{Edd} = 1.25 \times 10^{38} \left(\frac{M_{BH}}{M_\odot}\right)~ \rm erg~s^{-1}
\end{equation}
For $M_{BH}~\sim $ 6.0$\times$10$^8$ $M_\odot$ derived in this study, we obtain the Eddington luminosity to be 
7.5$\times$10$^{46}$ erg~s$^{-1}$ which is much larger than the jet power estimated above. This suggests that 
the jet of 1ES 1218+304 emits at sub-Eddington limit which is consistent with the broadband emission from 
the BL Lac type of blazars.
\subsection{1ES 1218+304: An Extreme Blazar}
According to the classification scheme based on synchrotron peak frequency in the broadband SED, 
blazars with synchrotron peak frequency above 10$^{15}$ Hz are referred to as 
\emph{High Synchrotron-Peaked} (HSP) sources \citep{Abdo2010}. Recent hard X-ray observations of 
HSP blazars indicate the presence of a small group of sources with synchrotron peak frequency above 
10$^{17}$ Hz \citep{Costamante2018,Costamante2002}. Other observational features of these sources include 
hard X-ray and TeV $\gamma$-ray spectra, large X-ray to radio flux ratio and lack of short term variability 
\citep{Bonnoli2015}. Sources with these peculiar characteristics are termed as \emph{Extremely High 
Synchrotron-Peaked} or simply \emph{Extreme Blazars} \citep{Singh2019}. Due to the hard X-ray spectrum ($\Gamma~ \le$ 2), 
the synchrotron peak in the low energy component of the SED is located at energies above few keV. 
Similarly, the intrinsic TeV spectrum of extreme blazars is exceptionally hard resulting which the 
IC peak in the broadband SED is located at energies above 1 TeV. Due to the hard intrinsic $\gamma$-ray 
spectrum, exterme blazars are relatively faint in the \emph{Fermi}-LAT energy range. In the present study, 
the time averaged X-ray spectrum of 1ES 1218+304 obtained from the long term \emph{Swift}-XRT observations is 
hard with photon spectral index of 1.99$\pm$0.01 (Section 4.3). Also, the intrinisic TeV spectra of the source 
derived from the MAGIC and VERITAS observations are described by a power law with photon spectral index $\le$ 2 (Section 4.4).
Therefore, the X-ray and TeV $\gamma$-ray spectral characteristics of 1ES 1218+304 indicate \emph{extreme} nature 
of the source. However, detection of the near simultaneous strong X-ray and TeV $\gamma$-ray flare from 1ES 1218+304 
on February 26, 2009 (MJD 54888) with variability at day timescale \citep{Archambault2014} contradicts the \emph{extreme} 
behaviour of the source. Therefore, 1ES 1218+304 is a normal HSP blazar and its extreme nature can be investigated in future 
from the dedicated long term simultaneous X-ray and $\gamma$-ray observations.
\section{Conclusions}
In the present work, we have reported the long term study of the blazar 1ES 1218+304 using 
10 years of contemporaneous MWL data from space and ground based observations 
between September 1, 2008 and August 31, 2018. This study provides first detailed investigation 
of the behaviour of broadband emissions from the source. The important findings about the nature of 
emissions from the blazar 1ES 1218+304 are :
\begin{itemize}
		
	\item Near simultaneous multi-wavelength light curves from radio to HE $\gamma$-ray indicate a
	      low emission state of the source. No significant change in the long term emission activity in 
	      different energy bands is observed.

       \item  The emissions in various wavebands exhibit intrinsic variability with small fractional variability 
	      amplitude ($F_{var}$ = 4.5-10$\%$) except in radio where variability is found to be relatively 
	      higher ($F_{var} \sim$ 18$\%$). However, no sginature of any short term variability is obtained 
	      in the MWL light curves.	
	
      \item  The X-ray emission in the energy range 0.3-10 keV is described by a hard power law with time averaged 
	      photon spectral index of $\sim$ 1.99. A hint for spectral hardening is observed in the scatter plot of 
		daily integral flux above 0.3 keV vs photon spectral index. 

      \item  The HE $\gamma$-ray emission in the energy range is also described by a power law with spectral index of 
		$\sim$ 1.67 suggesting that the blazar 1ES 1218+304 is a hard $\gamma$-ray source. The VHE $\gamma$-ray 
		emissions in the low activity state of this source detected by the MAGIC and VERITAS telescopes are 
		observed to suffer a large attenuation due to EBL absorption and the intrinsic VHE emission is consistent 
		with the hard photon spectral index of $\le$ 2.
      
      \item The broadband SED of the source can be broadly reproduced by a leptonic simple one zone SSC model with the electron 
	      energy distribution described by a log-parabolic function. The optical and UV emissions from the source are found to 
		be dominated by the stellar thermal emissions from the host galaxy and can be modelled using the \emph{PEGASE.3} code for 
		the evolution of galaxies and also by a simple black-body approximation. However, dedicated optical/UV observations of the 
		blazar 1ES 1218+304 are required to estimate the exact host galaxy contribution to the jet emission.	

     \item  The X-ray observations from the \emph{Swift}-XRT suggest that the mass of SMBH at the center of host galaxy is  		
            $M_{BH}~ \sim$ 6.0$\times$10$^8$ $M_\odot$ and the broadband emissions from the source are consistent with the 
           sub-Eddington jet activity.

    \item  Due to its hard X-ray and TeV $\gamma$-ray spectra, the blazar 1ES 1218+304 is an important source for probing (i) the 
	   particle acceleration mechanisms that can produce hard power law distribution in the jet and (ii) the intergalactic 
	   magnetic field using the cascade emission and (iii) cosmic infrared component of EBL. The source is also 
	 important for observations with the upcoming CTA (Cherenkov Telescope Array) observatory to establish the link between 
	 the TeV $\gamma$-ray emission and MeV-GeV emission measured from the \emph{Fermi}-LAT and its extreme blazar behaviour.
\end{itemize}
\section*{Acknowledgements}
Authors thank the anonymous reviewer for the critical and valuable comments to improve the manuscript. 
We acknowledge the use of public data obtained through \emph{Fermi} Science Support Center (FSSC) provided by NASA and 
\emph{Swift}-BAT transient monitor results provided by the Swift/BAT team. This research has made use of MAXI data provided by 
RIKEN, JAXA and the MAXI team. We also acknowledge the use of X-ray data supplied by the UK \emph{Swift} Science Data Centre at 
the University of Leicester. This research has made use of data from the OVRO 40-m monitoring program 
(Richards, J. L. et al. 2011, ApJS, 194, 29) which is supported in part by NASA grants NNX08AW31G, NNX11A043G, and NNX14AQ89G and 
NSF grants AST-0808050 and AST-1109911.

\bibliographystyle{mnras}
\bibliography{MS} 

\bsp	
\label{lastpage}
\end{document}